\documentclass[9pt]{article} 
\usepackage{spconf,amsmath,graphicx,cite}
\usepackage{bbm,amssymb}
\usepackage{pgfplots}
\usepgfplotslibrary{groupplots}
\usepackage{tikz}
\usetikzlibrary{plotmarks}
\usetikzlibrary{calc}
\usetikzlibrary{patterns}
\usetikzlibrary{shapes,arrows,snakes,backgrounds}
\usepackage{amsthm, amssymb,mathtools}
\usepackage{array}
\usepackage{mdwmath}
\usepackage[english]{babel}
\usepackage{multirow}
\usepackage{multicol}

\usepackage{standalone}
\usepackage{caption}
\usepackage{subcaption}
\newcommand{\argmin}{\operatornamewithlimits{argmin }}

\newcommand{\argmax}{\operatornamewithlimits{argmax}}

\title{Sparse Ternary Codes for similarity search \\ have higher coding gain than dense binary codes}
\name{Sohrab Ferdowsi, Slava Voloshynovskiy, Dimche Kostadinov, Taras Holotyak}
\address{Department of Computer Science, University of Geneva, Switzerland\\
$\lbrace$sohrab.ferdowsi, svolos, dimche.kostadinov, taras.holotyak$\rbrace$@unige.ch
}
\begin{document}

\maketitle
\begin{abstract}
This paper addresses the problem of Approximate Nearest Neighbor (ANN) search in pattern recognition where feature vectors in a database are encoded as compact codes in order to speed-up the similarity search in large-scale databases. Considering the ANN problem from an information-theoretic perspective, we interpret it as an encoding, which maps the original feature vectors to a less entropic sparse representation while requiring them to be as informative as possible. We then define the coding gain for ANN search using information-theoretic measures. We next show that the classical approach to this problem, which consists of binarization of the projected vectors is sub-optimal. Instead, a properly designed ternary encoding achieves higher coding gains and lower complexity. 
\end{abstract}
\section{Introduction}
\label{sec:intro}

The problem of content identification, e.g., identification of people from their biometrics or objects from unclonable features was first formalized in information-theoretic terms by Willems et. al. in \cite{1228096}. They defined the identification capacity as the exponent of the number of database items $M$ that could reliably be identified in an asymptotic case where the feature dimension $n \!\! \rightarrow \!\! \infty$. They modeled the enrollment and acquisition systems as noisy communication channels while they considered the vectors as random channel codes. The authors then characterized the identification capacity as $I(F;Q)$, the mutual information between the enrolled items $F$ and the noisy queries $Q$. In this setup, however, the increase in $n$ leads to an exponential increase in $M \simeq 2^{nI(F;Q)}$. This incurs infeasible search/memory complexities, making the system impractical. Subsequent works attempted at decreasing these complexities. For example, \cite{5205870} considered a two-stage clustering-based system to speed-up the search, while \cite{4839033} considered the compression of vectors prior to enrollment and studied the achievable storage and identification rates.

Similar to the content identification problem, the pattern recognition community considers the similarity search based on Nearest Neighbors (NN's) to a given query within the items in a database. This problem is the basis for many applications like similar image retrieval, copy detection, copywrite protection, etc. The idea is that semantic similarity can be mapped to distances within a vectorial space $\mathbb{R}^n$. The search for similarity then reduces to search for NN's by comparing the vectorial representation of a given query to representations of items in a database. The challenge, however, is when $M$ is huge, e.g., billions. The NN search based on linear scan then becomes the \textbf{main bottleneck} for the system. Approximative solutions, where performance is traded with memory and complexity, are then to be preferred. This is addressed in Approximate Nearest Neighbor (ANN) search, an active topic in computer vision and machine learning communities.\footnote{Refer to \cite{DBLP:journals/corr/WangSSJ14} for detailed review of ANN methods and applications.}

The methods introduced for ANN in pattern recognition communities can roughly be divided into two main categories. A first category of methods is based on quantization of the data. These methods mainly target memory constraints where every database item is given an index, or a set of indices which refer to codeword(s) from a trained codebook. At the query time, the items are reconstructed from the codebooks and matched with the given query, usually using look-up-tables.

A second family of methods for ANN, also considered in this work, is based on projecting the data from $\mathbb{R}^n$ to a usually lower dimensional space. The projected data are further binarized using the sign function. The search in the binary space is faster since it is less entropic than the original $\mathbb{R}^n$. Moreover, this approach brings more practical advantage since the distance matching in the binary space using the Hamming distance is performed in fixed-point instead of the floating point operations in the original real space or the space of codebooks in quantization-based methods.   

While the content identification literature is rigorously studying the fundamental limits of identification under storage or complexity constraints for known source and channel distributions and infinite code-lengths, the literature of ANN search drops these information-theoretic assumptions and hence treats the problem from a more practical perspective. This work tries to bridge these approaches. While we do not assume the asymptotic case and hence do not consider fundamental limits with their achievability and converse arguments, we propose a practical systems where it is explicitly required to maximize the mutual information between the query and database codes. Moreover, to address complexity requirements, we further require the codes to be sparse and hence allowing efficient search and storage. Concretely, this work brings the following contributions:
\begin{itemize}
\item Considering the projection approach to ANN, we introduce the concept of ``coding gain'' to quantify the efficiency of a coding scheme for similarity search. 

\item We then show that the standard approach in the literature, which is based on binarization of the projected values, is sub-optimal. Instead, as was also shown recently in \cite{Sohrab_WIFS2016}, the ternarization approach namely the Sparse Ternary Codes (STC) is to be preferred. While in \cite{Sohrab_WIFS2016} this advantage was shown from the signal approximation points of view, here we characterize it in terms of coding gain.


\item We next show how to design the STC codes to maximize the coding gain. As opposed to the Maximum Likelihood (ML) decoder, we consider a sub-optimal but fast decoder, which brings sub-linear search complexity. Considering the memory-complexity trade-offs, this provides a wealth of possibilities for design, much richer than the binary counterpart while maintains all its advantages. Based on this decoder, we simulate an identification scenario showing the efficiency of the proposed methodology.
\end{itemize}

The problem formulation, along with the definition of the proposed coding gain are discussed in section \ref{preliminaries}. Binary codes are summarized in section \ref{binary_codes} while the ternary encoding is discussed in section \ref{STC}. Comparison of the two encoding schemes in terms of the coding gain and also complexity ratio is performed in section \ref{Results}. Finally, section \ref{summary} concludes the paper.

\section{Preliminaries} \label{preliminaries}

\subsection{Problem formulation} \label{sub:Problem_formulation}
We consider a database $\mathcal{F} \!\! = \!\! \lbrace \mathbf{f}(1), \cdots, \mathbf{f}(M)\rbrace$ of $M$ items as vectorial data-points $\mathbf{f}(i)'$s, where $\mathbf{f}(i) = [f_1(i), \cdots, f_n(i)]^T$  $\in \mathbb{R}^n$ and $1 \leqslant i \leqslant M$. We assume, for the sake of analysis, that each database entry $\mathbf{f}(i)$ is a realization of a random vector $\mathbf{F}$ whose $n$ elements are \textit{i.i.d.} realizations of a random variable $F$ with $F \sim \mathcal{N}(0,\sigma_F^2)$.

The query $\mathbf{q} = [q_1, \cdots, q_n]^T \in \mathbb{R}^n$ is a perturbed version of an item from $\mathcal{F}$. We assume the perturbation follows an AWGN model as $\mathbf{Q} = \mathbf{F} + \mathbf{P}$, where elements of $\mathbf{P} \sim \mathcal{N}(\mathbf{0},\sigma_P^2 \mathbf{I}_n )$ are independent from elements of $\mathbf{F}$. For a given query, the similarity search is based on a distance measure $\mathcal{D}(\cdot,\cdot): \mathbb{R}^n  \! \times \! \mathbb{R}^n \rightarrow \mathbb{R}^+$, which is assumed here to be the $\ell_2$ distance, i.e., $\mathcal{D}_{\ell_2}(\mathbf{q},\mathbf{f}(i)) = ||\mathbf{q} - \mathbf{f}(i)||_2$.

In the case of the content identification problem, the search system finds the nearest item from $\mathcal{F}$ to $\mathbf{q}$, i.e., $\hat{i} \!\! = \!\! \argmin_{1 \leqslant i \leqslant M}{[\mathcal{D}_{\ell_2}(\mathbf{q},\mathbf{f}(i))]}$, or produces a list $\mathcal{L}(\mathbf{q})$ of nearest items in the more general (A)NN search, i.e., $\mathcal{L}(\mathbf{q})  = \lbrace i :  \mathcal{D}_{\ell_2}(\mathbf{q},\mathbf{f}(i))  \leqslant \epsilon n \rbrace$, where $\epsilon > 0$ is a threshold.
\subsection{Coding gain for similarity search} \label{sub:coding_gain}

The above problem formulation implies a memoryless  observation channel for queries. In \cite{1228096} the identification capacity is defined for this channel (along with another AWGN for the enrollment channel) as a measure for the number of items that can be reliably identified. While this analysis is based on the asymptotic assumption where the number of items $M$ should grow exponentially with dimension $n$, it can be argued, however, that $M$ is fixed in practice and can be even well below the amount that the capacity would accommodate. Moreover, the probability of correct identification might be only asymptotically achievable and is less than 1 in practice.

Therefore, in this work, instead of the channel coding arguments, we consider the practical case where the focus is on fast decoding for which the given database is encoded. This comes with the price of lower identification performance. 

To quantify the efficiency of a coding scheme for fast ANN search under the setup of section \ref{sub:Problem_formulation}, we consider the coding gain as the ratio of mutual information between the encoded versions of the enrolled items and query and the entropy of the enrolled items, i.e.: 

\begin{equation} \label{eq:coding_gain}
g_{\mathcal{F}}(\psi_e,\psi_i) = \frac{I(X;Y)}{H(X)},
\end{equation}
where the encoding for enrollment provides $X = \psi_e[F]$ and encoding for identification provides $Y = \psi_i[Q]$.\footnote{Notice that, unlike the usual binary design, in the STC framework, these two stages need not be the same. In fact, they are designed according to the statistics of the observation channel.}

   \begin{figure}  [!h]
\vspace{-0.75cm}   
   \begin{center}
   \begin{tabular}{c}
    \includegraphics[width=0.50\textwidth]{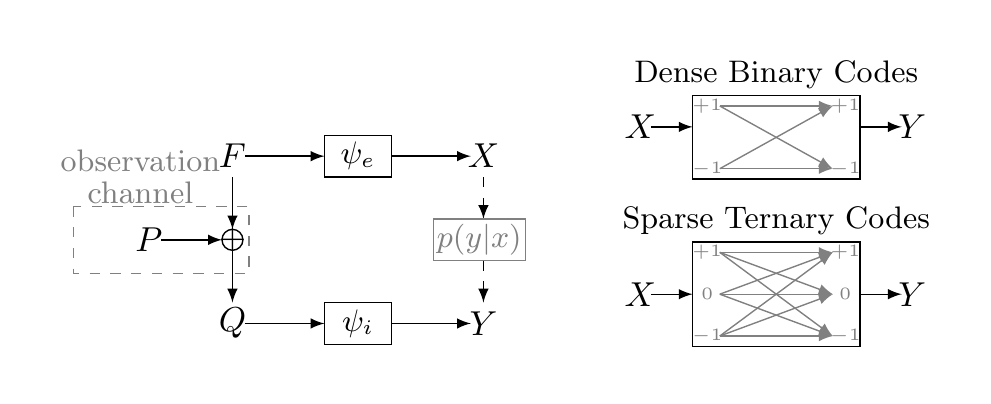}
   \end{tabular}
   \end{center}
   \vspace{-1.25cm}
\caption[example]{Encoders $\psi_e(\cdot)$ and $\psi_i(\cdot)$ map the feature space to the space of codes for enrollment and identification, respectively. A BSC models the perturbation in binary codes, while a noise-adaptive ternary channel models the STC.}
{ \label{fig:Channel_diagram}}
   \end{figure}

Mutual information in the definition of Eq. \ref{eq:coding_gain} can directly be linked to channel transition probabilities. The choice of entropy, on the other hand, is justified by both memory and complexity requirements. Obviously, the cost of the database storage is directly linked to $H(X)$ when source coding is used. Moreover, since the effective space size is $|\mathcal{X}^n| \approx 2^{nH(X)}$, a lower entropic space also implies a lower search complexity.

\section{Binary codes} \label{binary_codes}
Binary encoding is a classical approach widely used in the pattern recognition literature to address fast search methods. As mentioned earlier, they are based on projecting the data to a lower-dimensional space and binarizing them.

\subsection{Random Projections} \label{sub:RP}
The projection of vectors of $\mathcal{F}$ and $\mathbf{q}$ in $\mathbb{R}^n$ is performed using random projections to $\mathbb{R}^l$.\footnote{The recent trend in pattern recognition is favoring projectors which are learned from the data. However, in practice, it turns out that the performance boost obtained from these methods is limited to low bit-rate regimes only. Moreover, they are not straightforward for analytic purposes. Therefore, here we opt for random projections.} This choice is justified by results like Johnson-Lindenstrauss lemma \cite{johnson1984extensions}, where, under certain conditions, the pair-wise distances in the projected domain are essentially preserved with probabilistic guarantees.

Random projections are usually performed using an $n \times l$ unit-norm matrix $\mathrm{W}$ whose elements are generated as \textit{i.i.d.} from a Gaussian distribution $W \sim \mathcal{N}(0,\frac{1}{n})$. In the case of binary codes, for memory constraints, usually $l < n$. For the ternary codes proposed in \cite{Sohrab_WIFS2016}, which we analyze next, however, since the entropy can be bounded by imposing sparsity, longer code lengths can be considered. In this case, the use of sparse random projections of \cite{li2006very} is justified, where performance guarantees require longer lengths. This is very useful since the sparsity in the projector matrix can be reduced to $\mathcal{O}(\frac{2nl}{s})$ instead of $\mathcal{O}(nl)$, where $s$ is the sparsity parameter as in 
$W \sim
\begin{cases}
   \pm \sqrt{\frac{s}{2n}},   &\text{w.p.       }  \frac{2}{s}, \\
   0,    &\text{w.p.       }  1- \frac{2}{s}. 
\end{cases}$

According to the data model assumed in section \ref{sub:Problem_formulation}, the elements of the projected data $\tilde{F}$ and $\tilde{Q}$ will also follow Gaussian distribution as $\tilde{F}\sim \mathcal{N}(0,\sigma_F^2)$ and $\tilde{Q}|\tilde{F} \sim \mathcal{N}(\tilde{F}, \sigma_P^2)$, where $\tilde{\mathbf{F}}^T = \mathbf{F}^T W$ and $\tilde{\mathbf{Q}}^T = \mathbf{Q}^T W$. This gives the joint-distribution of the projected data as a bivariate Gaussian with $\rho = \frac{\sigma_F}{\sqrt{\sigma_F^2 + \sigma_P^2}}$, i.e.,  $p(\tilde{f},\tilde{q}) = \mathcal{N}([0,0]^T,\begin{bmatrix} \sigma_F^2 & \sigma_F^2\\ \sigma_F^2 & \sigma_F^2 + \sigma_P^2 \end{bmatrix})$.
\subsection{Binarizing the projections} \label{sub:binarization}
Binary codes are obtained by binarizing the projected values using the sign function, i.e.,  $X_b = \! \text{sign} (\tilde{F})$ and $Y_b = \text{sign} (\tilde{Q})$. One can consider an equivalent to the observation channel in the encoded case. In \cite{5592809}, for the \textit{i.i.d.} Gaussian setup, this channel was derived as a BSC with $I(X_b,Y_b) = 1 - H_2(P_b)$, where $P_b = \mathbb{E}_{p(\tilde{q})}[\mathcal{Q}(\frac{|\tilde{q}|}{\sigma_P})] = \!\! \frac{1}{\pi} \arccos(\rho)$ is the probability of bit flip and $\mathcal{Q}(u) = \!\! \int_u^{\infty} \!\! \frac{1}{\sqrt{2\pi}}e^{\frac{-u'}{2}} du'$ is the Q-function.

The maximum-likelihood optimal decoder for the binary codes is simply the minimum Hamming-distance decoder, i.e., $\hat{i} = \underset{{1 \leqslant i \leqslant M}}{\text{argmin}} [\mathcal{D}_{H}(\mathbf{y}_b,\mathbf{x}_b(i))]$ with $\mathcal{D}_H (\mathbf{y}, \mathbf{x}) = \frac{1}{l} \sum_{j = 1}^l y_j \oplus x_j$ and $\oplus$ representing the XOR operator. As mentioned earlier, this has two main practical advantages. First, the cost of storage of any database vector will reduce to $l$ bits. Second, the search is performed on the binary vectors using fixed-point operations. However, as we will show in the next section, better performance can be achieved with STC, the ternary counterpart of the binary codes.  

\section{Sparse Ternary Codes (STC)} \label{STC}
The idea behind the STC \cite{Sohrab_WIFS2016} is that different projection values have different robustness to noise which can be expressed as a bit reliability measure. This can be achieved by ignoring the values whose magnitude is below a threshold, i.e., $X_t = \phi_{\lambda_X}(\tilde{F})$ and $Y_t = \phi_{\lambda_Y}(\tilde{Q})$, where: 
\[
\phi_{\lambda}(t) =
\begin{cases}
   +1, & t \geqslant \lambda, \\
   0, & -\lambda <t < \lambda, \\
    -1, &  t \leqslant -\lambda, 
\end{cases}
\]
and $\lambda_X$ and $\lambda_Y$ are threshold values for the enrolled items and the query, respectively. The threshold values $\lambda_X$ and $\lambda_Y$ directly influence the sparsity ratios $\alpha$, the ratio of the number of elements of $\mathbf{X}_t$ with non-zero values ($\lbrace \pm 1 \rbrace$'s)  to the code length $l$ and $\gamma$, the ratio of the number of elements of $\mathbf{Y}_t$ with non-zero values to the code length $l$. These values can be easily calculated as:
\begin{align} 
\begin{split} 
\alpha &=  \int_{-\infty}^{-\lambda_X} p(\tilde{f}) d\tilde{f} + \int_{\lambda_X}^{\infty} p(\tilde{f}) d\tilde{f} = 2\mathcal{Q} \Big(\frac{\lambda_X}{\sigma_F} \Big), \\ 
\gamma &=  \int_{-\infty}^{-\lambda_Y} p(\tilde{q}) d\tilde{q} + \int_{\lambda_Y}^{\infty} p(\tilde{q}) d\tilde{q} = 2\mathcal{Q} \Bigg(\frac{\lambda_Y}{\sqrt{\sigma_F^2 + \sigma_P^2}} \Bigg).
\end{split} \label{eq:alpha_gamma}
\end{align}
Complementary to \cite{Sohrab_WIFS2016} where the choice of $\phi_{\lambda}$ was justified as an approximation for the hard-thresholding function, the optimal solution to the direct sparse approximation problem, here we propose an information-theoretic argument.

\subsection{Information measures for STC}
It is very useful to study the equivalent ternary channel between $X_t$ and $Y_t$. The element-wise entropy of the ternary code is:
\begin{equation} \label{eq:H_ternary}
H(X_t) = -2\alpha log_2(\alpha) - (1-2\alpha)log_2(1-2\alpha).
\end{equation}

According to Eq. \ref{eq:alpha_gamma}, this is only a function of the threshold value $\lambda_X$ (see Fig. \ref{fig:Gain_comparison}.a).



As for $I(X_t;Y_t)$, unlike the binary case we cannot have a closed form expression. Instead, we calculate this quantity with numerical integration. 

Consider the transition probabilities for ternary channel:
\begin{equation}
 \label{eq:transition_prob} 
\mathrm{P} = p(y|x)=
  \begin{bmatrix}
    p_{+1|+1} & p_{0|+1} & p_{-1|+1} \\
    p_{+1|0} & p_{0|0} & p_{-1|0} \\
    p_{+1|-1} & p_{0|-1} & p_{-1|-1} \\
  \end{bmatrix}.
\end{equation}
The elements of the transition matrix $\mathrm{P}$ are defined as the integration of the conditional distribution $p(\tilde{Q}|\tilde{F}) = \frac{p(\tilde{F},\tilde{Q})}{p(\tilde{F})}  $ with proper integral limits and calculated numerically for these threshold values. For example, $p_{+1|+1} = p_{\tilde{Q}|\tilde{F}}(+1|+1) = \frac{\int_{\lambda_Y}^{\infty} \int_{\lambda_X}^{\infty} p(\tilde{f},\tilde{q}) d\tilde{f} d\tilde{q}} {\int_{\lambda_X}^{\infty} p(\tilde{f}) d\tilde{f}}$ is the probability that values of $\tilde{F}$ greater than $\lambda_X$ are queried with values greater than $\lambda_Y$ while $p_{0|+1} = p_{\tilde{Q}|\tilde{F}}(0|+1) = \frac{\int_{-\lambda_Y}^{\lambda_Y} \int_{\lambda_X}^{\infty} p(\tilde{f},\tilde{q}) d\tilde{f} d\tilde{q}} {\int_{\lambda_X}^{\infty} p(\tilde{f}) d\tilde{f}}$ quantifies the probability that these values have magnitudes less than $\lambda_Y$ in the query and hence will be assigned as `0' in the ternary representation. Out of these 9 transition probabilities, 5 are independent and the rest are replicated due to symmetry.

The mutual information in the ternary case is:
\begin{equation*}
I(X_t;Y_t) = H(X_t) + H(Y_t) - H(X_t,Y_t),
\end{equation*}
where $H(X_t)$ is given by Eq. \ref{eq:H_ternary}. Similarly, $H(Y_t) = -2\gamma log_2(\gamma) - (1-2\gamma)log_2(1-2\gamma)$, where $\gamma$ can be derived from its definition of Eq. \ref{eq:alpha_gamma} or, equivalently $\gamma = \alpha (\mathrm{P}(1,1) + \mathrm{P}(1,3)) + (1-2 \alpha) \mathrm{P}(1,2)$. The joint entropy is then calculated from elements of $P$ as:

\begin{align} \label{eq:MI_STC}
\hspace{-0.5cm}
\begin{split}
H(X_t,Y_t) &=  -2 \alpha \mathrm{P}(1,1)log(\alpha \mathrm{P}(1,1)) \\
&-2 \alpha \mathrm{P}(1,2)log(\alpha \mathrm{P}(1,2)) -2 \alpha \mathrm{P}(1,3)log(\alpha \mathrm{P}(1,3))\\ 
&-2 (1-2\alpha) \mathrm{P}(2,1)log((1-2\alpha) \mathrm{P}(2,1))\\
& -(1-2\alpha) \mathrm{P}(2,2)log((1-2\alpha) \mathrm{P}(2,2)).
\end{split}
\end{align}
This, in fact, is a function only of $\lambda_X$, $\lambda_Y$ and $\sigma_P^2$. We will use these quantities in section \ref{sub:comparison} to compare the coding gain of these ternary codes with those of binary codes.  
%

\subsection{Fast sub-linear decoder} \label{sub:sublinear_dec}

\sloppy Given a ternary sequence $\mathbf{y} = (y_1,\cdots,y_l)^T$, the maximum-likelihood rule to choose the optimal $\mathbf{x}(i)$ amongst the registered $\mathbf{x}(1),\cdots,\mathbf{x}(M)$ is given as $\hat{i} = \argmax_{1 \leqslant i \leqslant M}  p_{\mathbf{Y}|\mathbf{X}}(\mathbf{y} |\mathbf{x}(i) )$. This rule can be simplified as the maximization of the log-likelihood as: 
\begin{equation}
\hat{i} = \argmax_{1 \leqslant i \leqslant M}  \sum_{j = 1}^l \log{p_{Y|X}(y_j|x_j(i))}.
\label{eq:ML_decoder}
\end{equation}
This requires to take into account all the 9 transition probabilities of Eq. \ref{eq:transition_prob}.

Similar to the binary codes, the above maximum-likelihood decoder should exhaustively scan all the $M$ items in the database. However, by skipping the transitions to and from `0' and only considering elements of $\mathrm{P}$ containing `+1' and `-1', an alternative decoder can be considered as:
\vspace{-0.5cm}

\begin{align} 
\begin{split}
\hat{i} = \argmax_{1 \leqslant i \leqslant M}  \sum_{j = 1}^n \Big[ \nu (\mathbbm{1}_{\lbrace x_j = y_j = +1 \rbrace} + \mathbbm{1}_{\lbrace x_j = y_j = -1 \rbrace}&) \\
+\nu' (\mathbbm{1}_{\lbrace x_j = +1, y_j = -1 \rbrace} + \mathbbm{1}_{\lbrace x_j = -1, y_j = +1 \rbrace}&) \Big],
\end{split}
\label{eq:sublinear_decoder_STC}
\end{align}
where $\nu$ and $\nu'$ are voting constants that will be specified shortly and $\mathbbm{1}_{\lbrace \cdot \rbrace}$ is the indicator function.

This decoder is sub-optimal in the ML sense since instead of taking into account all the 9 transitions (from which 5 independent), it considers only 4 transitions (from which 3 independent) and treats the other 5 as the same. However, it performs the search in sub-linear complexity instead of the otherwise exhaustive scan. As we will show next, this allows us to drastically reduce the search complexity. An algorithmic description of this decoder was given in \cite{Sohrab_WIFS2016} for practical scenarios. In this work, however, since we are assuming a known distribution for the data, we can calculate the optimal values of $\nu$ and $\nu'$ as Eq. \ref{eq:penalty}, where $\nu$ encourages sign match and $\nu'$ is a negative value that penalizes sign mismatch. Since we are ignoring the `0' transitions, $\nu^0$ is a bias term and compensates by considering the expectation of the occurrences of $\log{p(y|x)}$ terms in Eq. \ref{eq:ML_decoder}, when either $x$ or $y$ is `0'.

\begin{align} 
\begin{split} 
\nu &=  \nu^0 + log(\mathrm{P}(1,1)), \\ 
\nu' &=  \nu^0 + log(\mathrm{P}(1,3)),\\
\nu^0 &=  -\big[ 2\alpha \mathrm{P}(1,2) \log{\mathrm{P}(1,2)} \\
&+ (1-2 \alpha) (2\mathrm{P}(2,1) \log{\mathrm{P}(2,1) + \mathrm{P}(2,2) \log{\mathrm{P}(2,2)}})  \big]. 
\end{split} \label{eq:penalty}
\end{align}

\section{Simulation results} \label{Results}
\subsection{Comparison of Sparse Ternary Codes with dense binary codes} \label{sub:comparison}
\begin{figure*}  
   \begin{center} 

\includegraphics[width=1.0\textwidth]{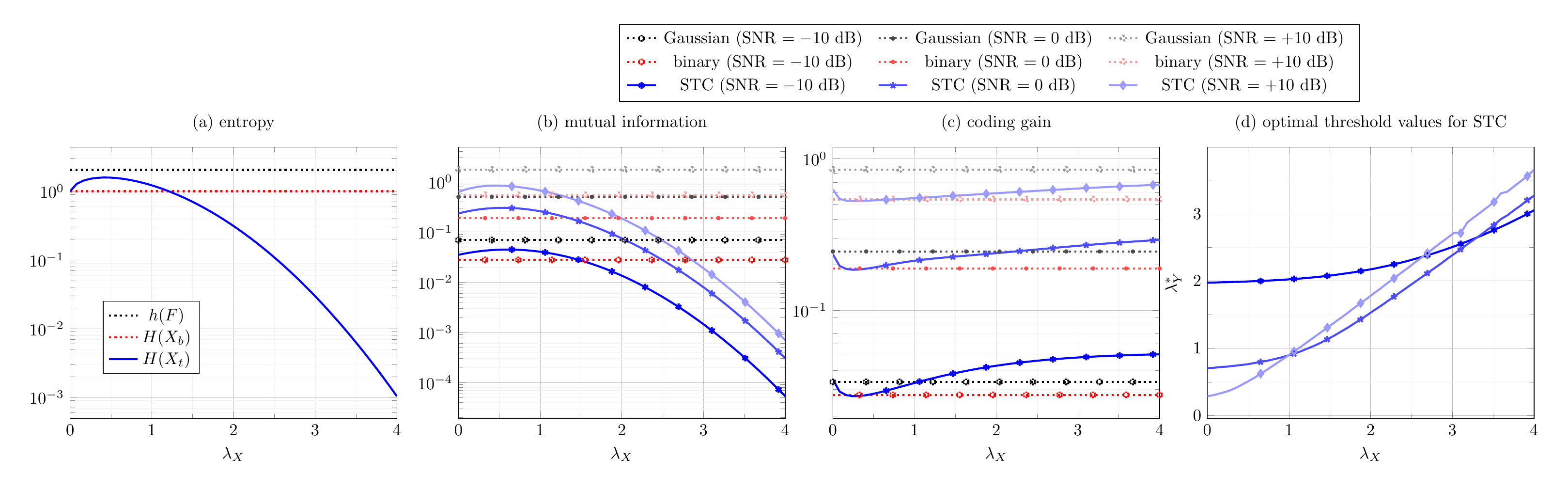}
   \end{center}
\vspace{-0.5cm}    
   \caption[example]{Coding gain comparison for ternary and binary encoding. Although the $I(X_t;Y_t)$ curve has a concave shape with respect to $\lambda_X$ which means that its value decreases as the code becomes sparser, $H(X_t)$ decreases with a faster rate for increasing $\lambda_X$. This means that the coding gain increases as the code become sparser and then saturates at $\frac{I(X_t;Y_t)}{H(X_t)} = 1 - \frac{H(X_t|Y_t)}{H(X_t)} \leqslant 1$. }
   \label{fig:Gain_comparison}
   \end{figure*}

Here we compare the coding gain of STC with dense binary codes. The mutual information of the ternary code depends on both $\lambda_X$ and $\lambda_Y$, therefore, for different values of $\lambda_X$\footnote{In practice, this is chosen based on memory constraints.}, we choose $\lambda_Y$ that maximizes $I(X_t;Y_t)$. Since there is no analytic expression for the maximum of $I(X_t;Y_t)$, for a fixed $\lambda_X$, we use a simple grid search among different values of $\lambda_Y$ for the maximizer of mutual information. Fig. \ref{fig:Gain_comparison}.d shows the ($\lambda_X, \lambda_Y^*$) pairs under three different noise levels, as characterized by SNR $= 10log_{10}\frac{\sigma_F^2}{\sigma_P^2}$.

For fair comparison, we choose the code lengths of binary and ternary cases ($l_b$ and $l_t$, respectively) to ensure the same code entropy, i.e., $l_bH(X_b) = l_t H(X_t)$. We then compare 
$l_bI(X_b,Y_b)$ with $l_tI(X_t,Y_t)$ in Fig. \ref{fig:Gain_comparison}.b.

%
%

 \begin{figure} 
   \begin{center} 
   \includegraphics[width=0.45\textwidth]{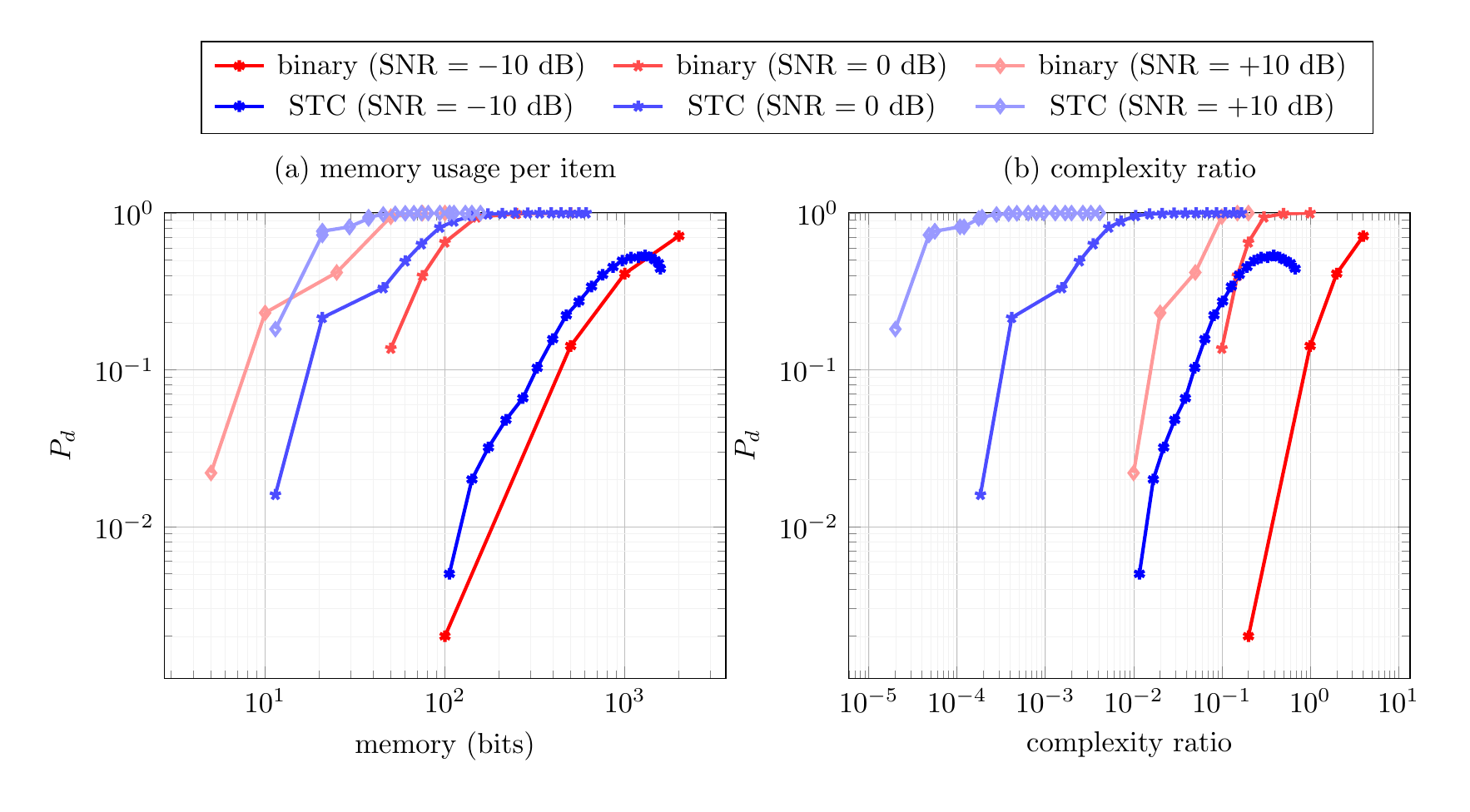}
   \end{center}
\vspace{-0.5cm}   
\caption[example]{Performance-memory-complexity profile for identification of $M = 1$ Mio synthetic data with $n = 500$. The sub-linear decoder of section \ref{sub:sublinear_dec} was used for STC.
}
   \label{fig:Identification} 
   \end{figure}

%
%

As is seen from this figure, the proper choice of thresholds leads to interesting regimes where, for the same entropy and hence the same number of bits, the ternary code preserves more mutual information compared to binary codes.


\subsection{Identification performance}
We consider the identification of synthetic data by comparing the probability of correct identification for different pairs of memory and complexity ratio in Fig. \ref{fig:Identification}. Memory usage is measured by entropy of a coded block, i.e., $l_bH(X_b)$ for the binary and $l_tH(X_t)$ and ternary while complexity ratio is measured as $\frac{l_b}{n}$ and $\frac{4\alpha \gamma l_t}{n}$, for the binary and ternary codes, respectively. Keeping the complexity of the sparse random projections stage the same in both cases in each experiment, for equal memory usage, a large gap is observed between the complexity ratios of the two counterparts. Furthermore, usually much better performance is achieved for the STC.


\vspace{-0.24cm}
\section{Conclusions} \label{summary}
This work is an attempt to bridge the problem of content identification with that of similarity search based on Nearest Neighbors in pattern recognition. Based on information-theoretic insights, the concept of coding gain was proposed as a figure of merit. It was shown that the Sparse Ternary Codes posses higher coding gain than the classical dense binary hashing scheme and hence provide better performance-complexity-memory trade-offs. As a future work, the extension to compression-based setups for the STC will be studied.



\bibliographystyle{IEEEbib}

\bibliography{tempBib}

\end{document}